\documentclass[oldversion]{aa}
\usepackage{txfonts}
\usepackage{graphicx}
\usepackage{natbib}
\bibpunct{(}{)}{;}{a}{}{,}
\usepackage{xspace}
\newcommand{\ascc} {\mbox{ASCC-2.5}\xspace}
\newcommand{\clucat} {\mbox{COCD}\xspace}

\newcommand{\olin}[1]{\overline{#1}}

\newcommand{\mc}[3]{\multicolumn{#1}{#2}{#3}}
\begin{document}

\title{The initial luminosity and mass functions of the Galactic open clusters}

\author{A.E.~Piskunov \inst{1,2,3}\and
        N.V.~Kharchenko \inst{1,3,4}\and
        E.~Schilbach \inst{3}\and
        S.~R\"{o}ser \inst{3}\and
        R.-D.~Scholz \inst{1}\and
        H. Zinnecker \inst{1} }

\offprints{R.-D.~Scholz}

\institute{Astrophysikalisches Institut Potsdam, An der Sternwarte 16, D--14482
Potsdam, Germany\\
email: apiskunov@aip.de, nkharchenko@aip.de, rdscholz@aip.de,
hzinnecker@aip.de
\and
Institute of Astronomy of the Russian Acad. Sci., 48 Pyatnitskaya
Str., 109017 Moscow, Russia\\
email: piskunov@inasan.rssi.ru
\and
Astronomisches Rechen-Institut, M\"{o}nchhofstra\ss{}e 12-14,
D--69120 Heidelberg, Germany\\
email: apiskunov@ari.uni-heidelberg.de, nkhar@ari.uni-heidelberg.de,
elena@ari.uni-heidelberg.de, roeser@ari.uni-heidelberg.de
\and
Main Astronomical Observatory, 27 Academica Zabolotnogo Str., 03680
Kiev, Ukraine\\
email: nkhar@mao.kiev.ua
}

\date{Received...; accepted ...}

\abstract{
We aim at the construction of luminosity and mass functions for Galactic open
clusters, based on integrated magnitudes and tidal masses. We also aim at
studying the evolution of these functions, with the ultimate purpose of deriving
the \textit{initial} luminosity and mass distributions of star clusters,
independent of model assumptions regarding the cluster mass-to-light ratio.
Finally we aim at a new determination of the percentage of field stars that have
originated in open clusters.

The integrated magnitudes are computed from individual photometry of cluster
members selected from the \ascc catalogue. The cluster masses we assumed to be
the estimated tidal mass recently published by us elsewhere. Analysis of the
cluster brightness distribution as a function of apparent integrated magnitudes
shows that the cluster sample drawn from the \ascc is complete down to apparent
integrated magnitude $I_V = 8$, with 440 clusters and compact associations above
this completeness limit. This, on average, corresponds to a completeness area in
the solar neighbourhood with an effective radius of about 1 kpc.

The observed luminosity function can be constructed in a range of absolute
integrated magnitudes $I_{M_V}= [-10,\,-0.5]$ mag, i.e. about 5 magnitudes
deeper than in the most nearby galaxies. It increases linearly from the
brightest limit to a turnover at about $I_{M_V}\approx-2.5$. The slope of this
linear portion is $a=0.41\pm0.01$, which agrees perfectly with the slope
deduced for star cluster observations in nearby galaxies. The masses of the
Galactic clusters span a range from a few $M_\odot$ to $\log M_c/M_\odot \approx
5.5$. The mass function of these clusters can be fit as a linear function with
log mass for $\log M_c/M_\odot > 2.5$, and shows a broad maximum between $\log
M_c/M_\odot=1.5$ and 2.5. For $\log M_c/M_\odot >2.5$, the linear part of the
upper cluster mass function has a slope $\alpha=2.03\pm0.05$, again in agreement
with data on extragalactic clusters. We regard this agreement as indirect
evidence that the tidal masses for Galactic clusters and the luminosity-based
masses for extragalactic clusters are on the same scale.

Considering the evolution of the cluster mass function now reveals a slight but
significant steepening of the slope with increasing age from
$\alpha=1.66\pm0.14$ at $\log t\leqslant6.9$ to $\alpha=2.13\pm0.08$ at $\log
t\leqslant8.5$. This indicates that open clusters are formed with a flatter
(initial) mass distribution than the overall observed cluster mass distribution
averaged over all ages. Interestingly, the luminosity function of open clusters
does not show the same systematic steepening with age as the mass function does.

We find that the initial mass function of open clusters (CIMF) has a two-segment
structure with the slopes $\alpha=1.66\pm0.14$ in the range $\log
M_c/M_\odot=3.37\dots4.93$ and $\alpha=0.82\pm0.14$ in the range $\log
M_c/M_\odot=1.7\dots3.37$. The average mass of open clusters at birth is
$4.5\cdot 10^3\,M_\odot$, which should be compared to the average observed mass
of about $700\,M_\odot$. The average cluster formation rate derived from the
comparison of initial and observed mass functions is
$\bar{\upsilon}=0.4\,\,\mathrm{kpc}^{-2}\mathrm{Myr}^{-1}$. Multiplying by the
age of the Galactic disc (T = 13 Gyr) the predicted surface density of Galactic
disc field stars originating from dissolved open clusters amounts to
$22\,M_\odot\,\mathrm{pc}^{-2}$ which is about 40\% of the total surface density
of the Galactic disc in the solar neighbourhood. Thus, we conclude that almost
half of all field stars were born in open clusters, a much higher fraction than
previously thought.

}


\maketitle

\section{Introduction}

Open clusters constitute an important part of a process transforming gas and
dust into stars. They are observed as the most prominent parts in the regions of
active star formation, or as tracers of the ceased star formation process in the
general Galactic field. However, the role they are playing in this process has
still not been fully understood. In spite of their prominence, there are indications
that classical open clusters contribute only 10\% or even less input
\citep{wiel71,misca78,clupop} to the total stellar population of the Galactic
disc. This contradiction can be explained either by an early decay of a
considerable fraction of newly formed star clusters \citep[see
e.g.][]{tut78,kah01,lamea} or by an insufficient knowledge of cluster formation
statistics. In this context, one should note that the most important items of
cluster formation like the distribution of cluster masses at birth (i.e., the
initial mass function of star clusters) and the cluster formation rate are still
poorly known.

Even in the close vicinity of the Sun, the only attempt to construct the
luminosity function of open clusters \citep{vdblaf84} is based on a sample of
142 clusters that, according to the authors, is to 2/3 complete within 400 pc.
A mass function of Galactic open clusters is not yet available.

In contrast, extragalactic cluster populations have been actively investigated,
especially during the present decade. The luminosity functions of remote clusters
are determined both in active and regular galaxies \citep[see e.g.][ and
references therein]{slars,grijs03,giel06}. For some galaxies, luminosity
distributions of open clusters are converted to the mass functions via the assumed
theoretical mass-luminosity relation \citep{zhang99}.

Using extragalactic data for the study of cluster population gives certain
advantages. When observing external galaxies, one is able to cover large areas
of their surfaces and catch the brightest objects, which are intrinsically rare.
This is rather difficult in the case of our own Galaxy since they could be
hidden by a dusty environment. Also, different selection effects can be 
taken into account better in an external galaxy because its clusters are located at
about the same distance from the Sun. Moreover, studies of extragalactic
clusters give a key to our understanding of the cluster formation in different
environments.

However, there are a number of problems related to studies based on
extragalactic clusters. Samples of extragalactic clusters are restricted to
relatively bright objects. There is a danger of a contamination of cluster
samples by brightest stars and of losing sparse objects in the strong
background. For example, \citet{slars} finds that his sample in NGC6949 is
contaminated to about 20\% by bright stars. The major problem is the necessity
of transforming the observed luminosities into cluster masses. The
corresponding transformations are mainly based on the luminosity of a dozen of
the most massive stars, whereas cluster masses are defined by numerous stars of
lower masses that are below the observing limit.

In our study of the local population of Galactic open clusters, we obtained
data that now allow a reliable construction of their luminosity and mass
functions. Our cluster sample contains 641 open clusters and 9 associations identified in the all-sky compiled catalogue \ascc \citep{kha01}.
For each cluster we determined a homogeneous set of cluster parameters
including, among others, its distance, age, and tidal mass. The sample is found
to be complete up to a distance of about 850~pc from the Sun \citep[]{clupop}.
This completeness limit corresponds to a distance modulus $(V-M_V)$ of about
10--10.5 \citep[]{clusim} and allow probes of the luminosity function of open
clusters that are much deeper than in other galaxies.

In this paper we aim at constructing the luminosity and mass functions of
the Galactic open clusters and at comparing them to the data in other
galaxies in order to touch the issue of universality of the cluster formation
process in different environments. We also consider temporal variations of
the two distributions. The ultimate purpose of this study is to construct the
initial luminosity and mass distributions of star clusters independent of model
assumptions on the M/L-ratio, to derive their parameters (the shape and limits),
and to draw conclusions on the impact of these results on our understanding of
the role of open clusters in the evolution of the stellar population of the
Galactic disc.

The paper has the following structure. In Sect.~\ref{sec:dadef} we briefly
describe the input data and give the definitions used throughout the paper.
Sect.~\ref{sec:lmf} is devoted to the construction of cluster the luminosity and
mass functions. In Sect.~\ref{sec:evol} we examine how cluster mass and
luminosity functions evolve with time and build the cluster initial mass
function. In Sects.~\ref{sec:discuss} and \ref{sec:concl} we discuss and
summarise the results.

\section{Data and definitions}\label{sec:dadef}

In this paper we use the results of the previous work based on the study of 650
nearby open clusters identified in the ASCC-2.5. The sample includes 520 known
objects and 130 open clusters detected within our project \citep[]{newclu}. For
each star projected on a cluster area, a membership probability was determined
in an iterative process that takes spatial, photometric, and proper
motion distributions of stars into account within the corresponding area on the sky
\citep[]{starcat}. At the end of the iterations for each cluster, we obtained
new coordinates of the cluster centre, the cluster size, the mean proper motion,
the distance from the Sun, reddening, and age \citep[][]{clucat,newclu}. These
parameters were determined with data on the most probable members, i.e., stars
having both kinematic and photometric membership probabilities higher than 61\%.
The results are included in the Catalogue of Open Cluster Data (COCD) and its
Extension \citep{clucat,newclu}.

In \citet{clumart} we published masses of 236 clusters of our sample estimated
from tidal radii determined with a three-parameter fit of King's profiles to the
observed density distributions of cluster members. To obtain mass
estimates for all clusters, we used these data to establish a relation between
tidal radius and the observed semi-major axis of the apparent
distribution of cluster members on the sky. The resulting tidal masses (i.e.
calibrated tidal masses) for 650 clusters have been published in \citet{clumart1}.
Though calibrated tidal masses of individual clusters are less accurate, their
distribution fits well the distribution of the tidal masses based on the direct
fitting of the King model. Moreover,  because the parameters are homogeneous and more 
numerous, the calibrated masses are better suited to statistical
investigations. In the following, we use the data on the calibrated tidal masses
to study the luminosity and mass functions of the Galactic open clusters.
Throughout the paper, masses are given in the units of solar masses.

Our list of 650 Galactic open clusters identified in the ASCC-2.5
includes  9 cluster-like associations (e.g. Vel~OB2, Sco~OB4), which require
further commentary. These are objects, which are classified both
as clusters and associations in the literature (e.g. $\alpha$ Per = Per OB3). They are included in
our sample because their observed properties (spatial structure, internal
kinematics, and stellar content) make them indistinguishable from regular open
clusters, although they have larger sizes and higher masses. Hereafter we refer to them
as cluster-like associations \citep[see also ][]{newrv07}.

We consider two young clusters, NGC~869 (also known as $h$~Per) and NGC~884
($\chi$~Per), as a single entity. They overlap in the projection on the
sky, share a huge corona, and have a large number (more than 50\%) of members in
common, making an accurate determination of their individual tidal radii and
masses rather difficult. Also, we exclude the cluster Mamajek~1 ($\eta$
Chamaeleontis) for which we were able to identify only three members in the \ascc.
According to Fig.~\ref{fig:agemass}, this cluster can be omitted without
consequences for the results. Therefore, our final cluster sample includes 648
entities.

As a measure of cluster brightness needed for the construction of the luminosity
function, we take the integrated magnitudes of the clusters. Doing so, we can
directly compare our results with findings based on observations of open
clusters in other galaxies. We define apparent and absolute integrated
magnitudes $I_V$ and $I_{M_V}$ as
\[
I_V=-2.5\,\log\left(\sum\limits_i^{N_1}\, 10^{-0.4\,V_i}+10^{-0.4\,\Delta
I_V}\right)\,,
\]
and
\[
I_{M_V} = I_V - (V-M_V)\,,
\]
where $N_1$ and $V_i$ are the number and the apparent magnitudes of the most
probable cluster members, respectively, and $(V-M_V)$ is the apparent distance
modulus available in the \clucat for each cluster. For ``unseen'' stars, $\Delta I_V$ is a magnitude
correction, introduced to make the computed $I_V$ and
$I_{M_V}$ independent of the range of stellar magnitudes actually observed in a
given cluster.

\begin{figure}
\resizebox{\hsize}{!}{
\includegraphics[clip=,angle=270]{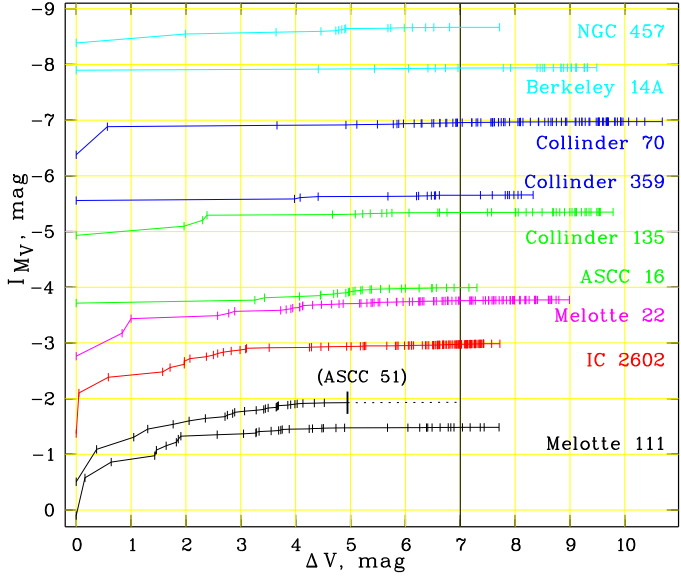}}
\caption{Absolute integrated magnitude $I_{M_V}$ profiles for 9 ``template''
clusters. $\Delta V$ is the magnitude difference between the brightest cluster
star and fainter members. The vertical bars indicate individual stars. The
contribution of stars fainter than $\Delta V=7$ to $I_{M_V}$ is small enough to be
neglected. As an illustration of the reduction method applied to the
observed integrated magnitudes of clusters, we show the profile of ASCC~51, which
ends at $\Delta V\approx 5$ mag (longer vertical bar). The extension (dashed
curve) to $\Delta V=7$ mag is based on the template Mel~111, and it results in
$\Delta I_V=0.007$ mag.
}
\label{fig:mag_int}
\end{figure}

To compute $\Delta I_V$, we used $(I_{M_V},\,\Delta V)$ profiles
of nine ``template'' clusters shown in Fig.~\ref{fig:mag_int}. The templates
were selected to satisfy two constraints: to be uniformly distributed over the
brightest stellar magnitudes of the clusters and to represent a sufficiently
long observed main sequence (MS). In Fig.~\ref{fig:mag_int} we show the
profiles as a function of $\Delta V_i=V_i-V_{br}$ ($V_i$ are individual
magnitudes of the most probable members of a template and
$V_{br}=\mathrm{min}\{V_i\}$). At $\Delta V_i=0$, the integrated magnitude
$I_{M_V}$ is identical to the absolute magnitude of the brightest cluster
member. As one can see, stars that are seven or more magnitudes fainter than the
brightest member ($\Delta V\geqslant7$) do not have a significant impact on
$I_{M_V}$. On the other hand, the typical absolute magnitude of cluster
stars at $\Delta V=7$ is brighter than $M_V=6$ or $m>0.9\,M_\odot$ in our
sample. In this mass range, the mass functions of cluster stars are only weakly
dependent on cluster age \citep{bamak03}. This enables us to safely apply the
above templates to the whole cluster sample.

A correction $\Delta I_V$ is computed for each cluster having a short MS as
\[
\Delta I_V = I_V^{fnt}-2.5\,\log\sum\limits_{V_i=V_{fnt}}^{V_{br}+\Delta
V}\,10^{-0.4\,V_i}
\]
after normalization of the suitable template profile
(see Fig.~\ref{fig:mag_int}).
Here $V_{br}$ and
$V_{fnt}$ are the brightest and faintest magnitudes of the most probable members
in the cluster under consideration, $V_i$ are template magnitudes, and $\Delta
V=7$~mag. The constant $I_V^{fnt}$ is computed assuming that the cluster and the
matching template have equal integrated magnitudes at $V_{fnt}$. Since the
integrated magnitude is mainly defined by the brightest cluster members, $\Delta
I_V$ is relatively small: it is always less than 0.85~mag, with an average of
0.13~mag, and for 95\% of the clusters $\Delta I_V<0.3$. On the other hand, in
33 clusters the MS exceeds a length of 7 magnitudes. We truncated their profiles
to $\Delta V=7$, for homogeneity.

The other membership samples, i.e., samples including stars with a membership
probability less than 61\%, produce somewhat brighter magnitudes (on average, by
about half a magnitude) but they are more strongly contaminated by field stars than
is the sample of the most probable cluster members.

Throughout the paper we use the following definitions of cluster luminosity and
mass functions. If the $\Delta N$ is the number of open clusters having absolute
integrated magnitude in the range $[I_{M_V}\,,I_{M_V}+\Delta I_{M_V}]$ and which
are observed in an area $\Delta S$ in the Galactic disc, then the luminosity
function $\phi$ is
\[
\phi(I_{M_V}) = \frac{1}{\Delta S}\frac{\Delta N}{\Delta I_{M_V}}\,;
\]
i.e., the luminosity function is a surface density distribution of open clusters
over the integrated magnitude. Similarly, the mass function $\eta$ is defined as
a surface density distribution of open clusters over the logarithm of mass:
\[
\eta(M_c)=\frac{1}{\Delta S}\frac{\Delta N}{\Delta\log M_c}\,.
\]
Since the above distributions include clusters of all ages, it is reasonable to
call $\phi(I_{M_V})$ and $\eta(M_c)$ the present-day luminosity and mass
functions of clusters, or CPDLF and CPDMF, respectively. In the following, we
also consider the luminosity and mass distributions of clusters confined by some
upper limit of their age $t$. We call them current luminosity/mass functions of
clusters or simply cluster luminosity/mass functions with the abbreviations CLF
and CMF. To distinguish them from present-day distributions we denote
them as $\phi_t(I_{M_V})$ and $\eta_t(M_c)$. Both functions are in fact cumulative with respect to age distributions.

The aim of the present paper is the construction of an initial mass function of
star clusters, which indeed describes the 
initial distribution, i.e. the distribution after re-virialisation 
after residual gas expulsion \citep{krobo02}. 
Hereafter, we
denote the initial luminosity/mass distributions as CILF and CIMF. A formal
definition of these functions is given in Sect.\ref{sec:evol}.

Finally, we must keep in mind the evolutionary status of open clusters included
in our sample. Since cluster membership is based on the proper motion data mainly
obtained in the optical spectral range, we consider our sample as representative
of optical clusters or ``classical'' open clusters. The embedded objects (e.g.,
the nearby cluster NGC~1333) are not included in our statistics since their
members would be fainter than the limiting magnitude of $V\approx11.5$ of the
\ascc if observed in the optical. Therefore, we assume the beginning of the
transparency phase after the removal of the bulk of the placental matter to be a
starting point of the evolution of a classical open cluster. The corresponding
age $t_0$ is defined by the lowest age of our clusters, that is, about 4 Myr.

\section{Magnitude-limited sample of open clusters and the observed luminosity
and mass functions}\label{sec:lmf}

\begin{figure}
\resizebox{0.95\hsize}{!}{
\includegraphics[angle=-90,clip=]{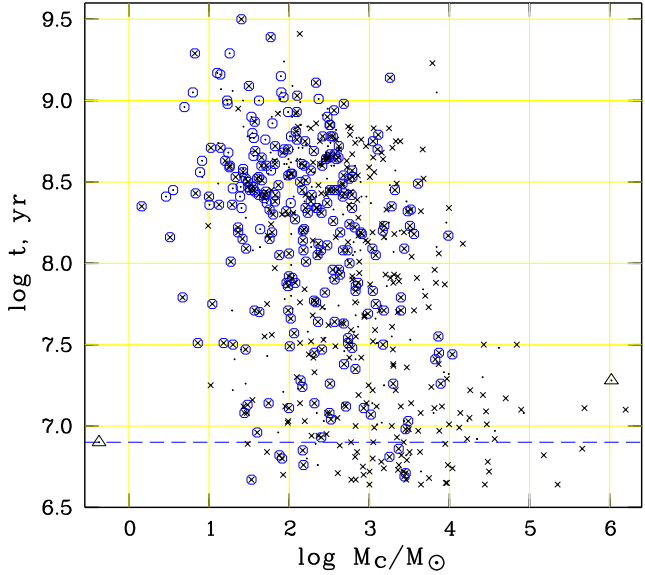}
}
\caption{Distribution of 650 clusters of our sample in $(\log t,\,\log
M_c)$-diagram (dots). Open circles mark volume-limited sample of 257 clusters with a
border of 850 pc. Crosses are 440 clusters from the magnitude-limited sample
constructed in Sect.~\ref{sec:lmf}. Triangles mark two clusters omitted from the
consideration (see Sect.~\ref{sec:dadef}). The dashed line shows the upper age
limit of the youngest subsample discussed in Sect.\ref{sec:evol}.
}
\label{fig:agemass}
\end{figure}

As shown in \citet{clupop}, our sample can be considered as volume-limited with
a boundary of about 850 pc. This volume contains 257\footnote{This number does
not include the two nearest open clusters UMa and the Hyades, for which we could
not determine cluster membership via the standard pipeline and, therefore, were
not able to obtain cluster masses and integrated magnitudes.} clusters, i.e.
about one third of the total sample. However, the completeness distance varies
if one considers clusters with different integrated magnitude. Very faint
clusters can possibly be seen within slightly smaller distances. In contrast,
very bright and massive clusters are seen at much greater distances but they are
rare, and the limit of 850~pc is not sufficient for a realistic estimate of
their average density. This presumption is supported by Fig.~\ref{fig:agemass}
where in an ``age-mass'' diagram we compare the distributions of all clusters
with the clusters belonging to the volume-limited sample. For $\log t < 7.5$,
the volume-limited sample does not represent the actual mass distribution of the
young clusters and cannot be used for a realistic estimate of a mass function.
The same is true for the distribution of integrated absolute magnitudes of young
clusters. The problem can be solved if we could extract a cluster sample that 
is limited by a certain integrated magnitude. In this case, we can determine the
completeness areas for clusters of different intrinsic brightnesses.

\begin{figure}
\resizebox{\hsize}{!}{
\includegraphics{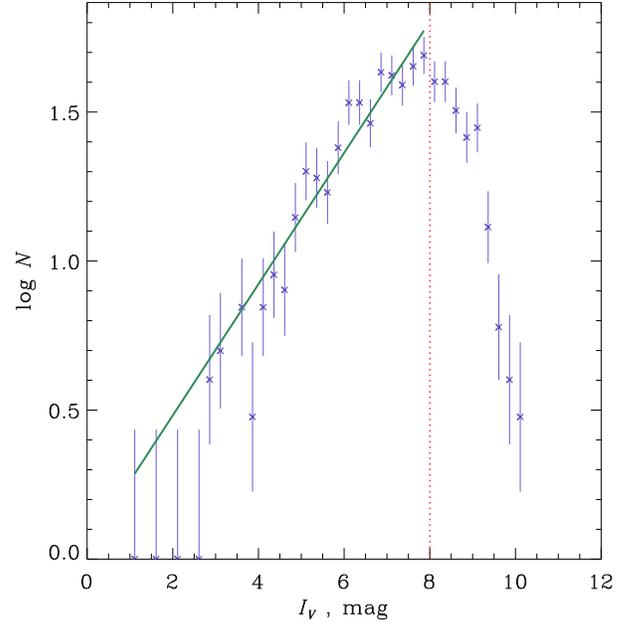}}
\caption{Distribution of open clusters versus apparent integrated magnitude
$I_V$ computed from data on the most probable cluster members (crosses). The
bars are corresponding Poisson errors, the solid line is a linear fit of the
distribution and the vertical dotted line is the adopted completeness limit
$\hat{I}_V$.
}
\label{fig:cmpl_mag}
\end{figure}

In Fig.~\ref{fig:cmpl_mag} we show the distribution of our clusters in bins of
apparent integrated magnitude $I_V$. Starting from $I_V \approx 3$, the
logarithm of cluster numbers is linearly increasing with increasing $I_V$ and
reaches a maximum at $\hat{I}_V=8$. We take $\hat{I}_V=8$ as the completeness
magnitude of our sample. In total, 440 clusters are brighter than $\hat{I}_V$.
This increases the basis for the following statistical study by a factor of
$\approx 1.7$ (compared to the previous number of 257).

The cluster luminosity function $\phi$ is constructed as a function of
integrated absolute magnitudes $I_{M_V,i}$ binned with a step size $\Delta
I_{M_V} = 0.25$, as the sum of partial surface densities
\[
\phi(I_{M_V,i}) =\frac{1}{\Delta
I_{M_V}}\sum\limits_j^{n_i}\frac{1}{\pi\,{\hat{d}_j}^{\,2}}\,.
\]
Here $\hat{d}_j$ is the individual completeness distance corresponding to
$\hat{I}_V=8$, $n_i$ is the number of clusters in the $i$th magnitude bin
$I_{M_V,i}$, and $\sum n_i=440$. Similarly, the mass function $\eta$ is computed
by using the logarithmic mass scale $\log M_{c,i}$ binned with a step size
$\Delta\log M_c=0.15$
\[
\eta(M_{c,i}) =\frac{1}{\Delta\log
M_c}\sum\limits_j^{\nu_i}\frac{1}{\pi\,{\hat{d}_j}^{\,2}}
\]
with $\nu_i$ the number of clusters in the mass bin $\log M_{c,i}$ and
$\sum \nu_i=440$. The completeness distance $\hat{d}_j$ for the $j$th cluster is
given as
\[
\log \hat{d}_j = 0.2\,[\hat{I}_V - I_{M_V,j} - 3.1\,E_j(B-V)] + 1\,,
\]
where $E_j(B-V)$ is the reddening for the cluster.

\begin{figure}
\resizebox{\hsize}{!}{
\includegraphics{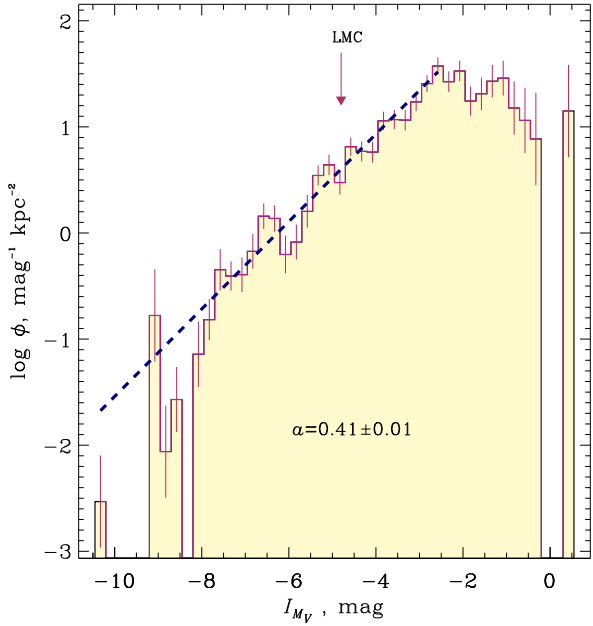}
}
\caption{Luminosity function of Galactic open clusters based on 440 local
clusters brighter than the completeness limit $\hat{I_V}$ of the sample. The
bars are Poisson errors, the dashed line shows a linear fit for the brighter
part of the histogram ($I_{M_V} < -2.5$) where $a$ is the corresponding slope.
The arrow indicates the limit of integrated absolute magnitudes reached for open
clusters in the LMC \citep[see][]{slars}.
}
\label{fig:histolf}
\end{figure}

In Fig.~\ref{fig:histolf} we show the present-day luminosity function CPDLF
constructed from 440 clusters. The CPDLF splits into a long quasi-linear portion
at bright magnitudes, a turnover between $I_{M_V}=-3$ and $-2$, and an apparent
decrease at fainter magnitudes. As expected, for Galactic star clusters the
CPDLF can be observed as much deeper than for clusters in other galaxies. For the
Large Magellanic Cloud (LMC), the faint limit is reached already at about
$I_{M_V}=-5$, and it is much brighter in more distant galaxies
\citep[][]{slars}.

Since the luminosity function is obtained from the distribution of clusters
within the respective completeness limit, the turnover can be considered as a
real feature of the CPDLF indicating a decrease of cluster density at $I_{M_V} >
-2.5$. The slope $a$ on the left side of the turnover ($I_{M_V} < -2.5$ ) is
computed by fitting to the histogram (Fig.~\ref{fig:histolf}) a function
\[
\log \phi = b + a\,I_{M_V}\,,
\]
where $a = 0.41\pm0.01,\,\,b=2.58\pm0.07$. This corresponds to a slope of the
cluster luminosity function $\alpha=2.02\pm0.02$ counted in luminosity intervals
(for a compatibility with stellar luminosity function we adopt that $dN/dL_V
\propto L^{-\alpha}$). The slope $\alpha$ is frequently used for the description
of the luminosity function of extragalactic clusters \citep[see e.g.][]{slars}
and can be expressed with the adopted definition as $\alpha= 2.5\,a+1$.

The only attempt to obtain the luminosity function of Galactic open clusters was
undertaken by \citet{vdblaf84}. These authors constructed the CPDLF from
integrated magnitudes computed by \citet{sagea83} for a sample of 142 clusters
with accurate $UBV$ photometry and proper motion membership. Their CPDLF was
obtained for $I_{M_V}\le -2$ and with a slope $a=0.2$~mag$^{-1}$, which is
flat compared to our result. The authors regard the sample to be 2/3 complete
within 400 pc and strongly incomplete outside this limit. However, no further
details are given about the completeness issue (e.g. how it depends on cluster
magnitude), and it is not possible to judge if their low value for the slope is
a consequence of the incompleteness.

\citet{slars} studied the luminosity functions of open clusters in six nearby
spiral galaxies constructed from HST archive images, and in the LMC from
literature data. Typically derived in the magnitude range $[-10,-7]$, the CPDLF
slope varies within $a\approx0.4\dots0.6$ mag$^{-1}$. This result coincides 
with our finding for open clusters in the Milky Way. Further evidence for an
agreement of the CPDLF slopes between Galactic and extragalactic clusters can be
found e.g., in \citet{grijs03}.

The other interesting feature in our CPDLF is a deficiency of clusters for
$I_{M_V}<-8$ compared to the fitted power law. A similar effect was observed by
\citet{slars} for extragalactic clusters.

\begin{figure}
\resizebox{\hsize}{!}{
\includegraphics{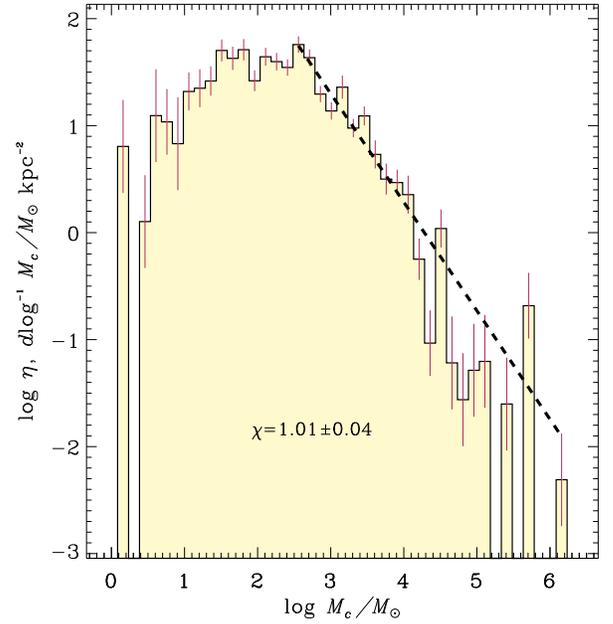}
}
\caption{Mass function of Galactic open clusters based on 440 local clusters
brighter than the completeness limit $\hat{I_V}$ of the sample. The bars are
Poisson errors, the dashed line shows a linear fit for the mass rich part of
the histogram ($\log M_c>2.5$) where $\chi$ is the corresponding slope.
}
\label{fig:histomf}
\end{figure}

The corresponding present-day mass function CPDMF of 440 clusters is shown in
Fig.~\ref{fig:histomf}. We note similar details in the CPDMF that we have just
observed for the CPDLF in Fig.\ref{fig:histomf}. One recognises a quasi-linear
portion in $\eta(M_c)$ for $\log M_c>2.5$, a turnover at about $\log M_c= 2$,
and a decrease for smaller masses. Again, we stress that the turnover of the
CPDMF can be considered as a real feature, since the determination is based on
unbiased cluster data. In the high-mass part of the histogram at $\log M_c>2.5$,
the CPDMF can be expressed in a canonical power-law form as
\begin{equation}
\log\eta = \log\eta^* - \chi\,\log M_c\,. \label{eq:eta}
\end{equation}
A fit of Eq.~\ref{eq:eta} to the observed mass function provides
$\chi=1.01\pm0.04$ for the slope and $\log\eta^*=4.3\pm0.14$ for the zero-point.
A deficiency of clusters at the high-mass end ($\log M_c\gtrapprox 4.0$) is less
pronounced than we observe for the luminosity function, and discrepancies to the
fitted relation are almost within Poisson errors.

Since all published data on mass functions of extragalactic clusters are based
on the mass-luminosity relation used to convert the observed photometric or
spectroscopic data into masses, only a formal comparison with our result is
possible. In \citet{grijs03}, the slopes of the CPDMFs are listed for open
clusters observed in four galaxies. Within a mass range $(10^3,10^6\,M_\odot)$,
a typical value of the CPDMF slope is $\alpha\approx2$. On the logarithmic
scale, this corresponds to $\chi\approx1$ and agrees well with our results. We
consider this result as indirect evidence of the coincidence of mass scales
of Galactic and extragalactic clusters.

\begin{figure}
\resizebox{\hsize}{!}{
\includegraphics{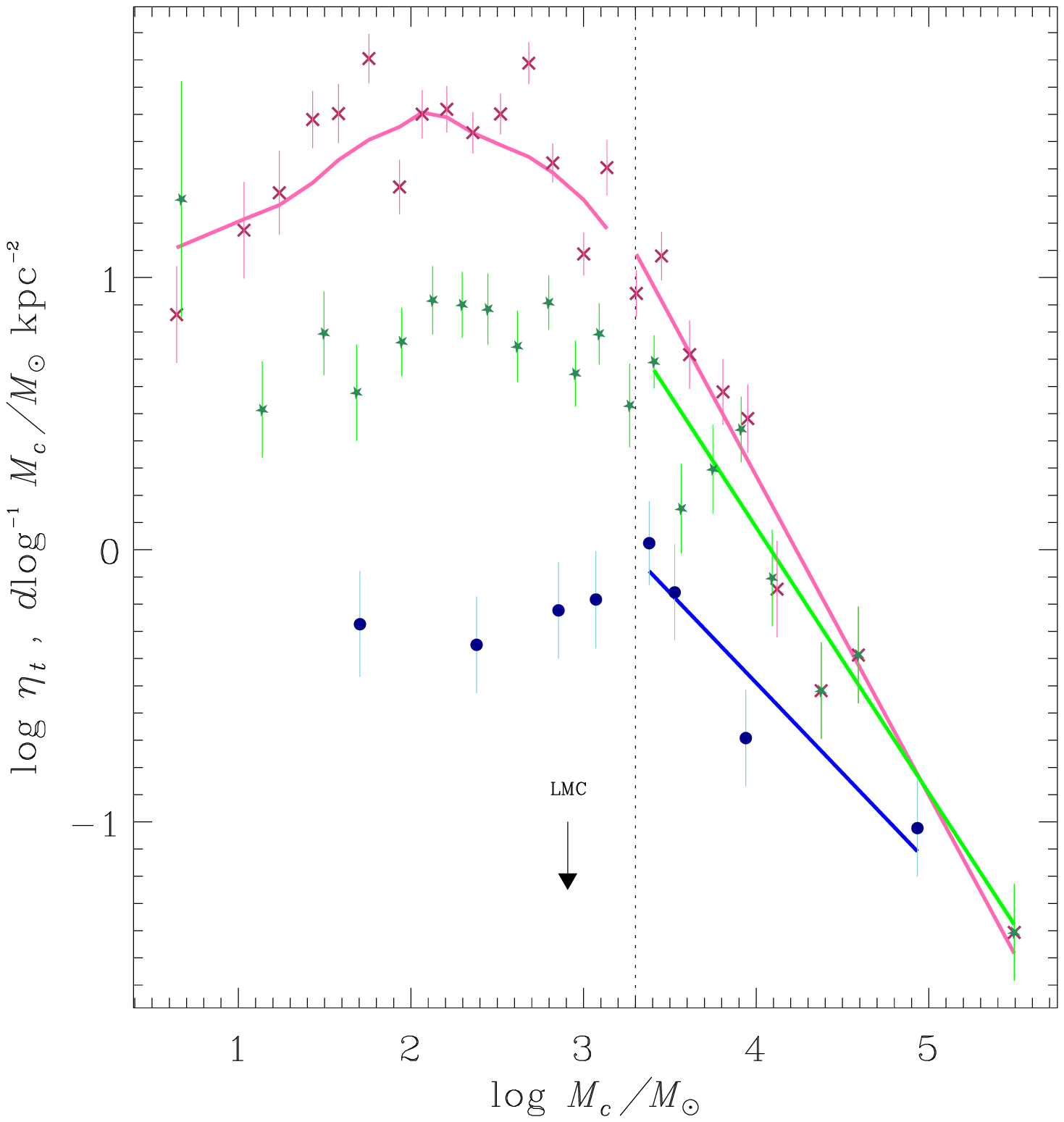}
}
\caption{Evolution of the mass function $\eta_t(\log M_c)$ of Galactic open
clusters. Different symbols mark samples with different upper limits of cluster
ages. The filled circles are for clusters with $\log t \leqslant 6.9$, stars
show CMF for $\log t \leqslant 7.9$, and crosses indicate the CPDMF based on all
440 clusters ($\log t \leqslant 9.5$). The bars are Poisson errors. The straight
lines are the corresponding fits to linear parts of the mass functions at masses
greater than $\log M_c=3.3$ indicated by the vertical dotted line. The curve is
the smoothed CPDMF at $\log M_c<3.3$. The arrow indicates the lower mass limit
reached for open clusters in the LMC.
}
\label{fig:fuma}
\end{figure}

\section{Evolution of luminosity and mass functions; cluster IMF}
\label{sec:evol}

Since star clusters evolve by changing the basic parameters (like mass and
integrated luminosity), their mass and luminosity functions also undergo
evolutionary changes. For stars, the impact of their evolution on the stellar
mass and luminosity functions was revealed for the first time by \citet{salp55},
who developed a receipt for the construction of stellar initial luminosity/mass
functions from present-day distributions.

Compared to stars, the case of stellar clusters is rather complicated. The
evolution of a cluster follows two independent time scales. The first one is the
nuclear time scale that governs the evolution of stars. The nuclear scale is
primarily responsible for changing the cluster's luminosity and is related to
the luminosity function of clusters. The second time scale is defined by the
dynamical evolution and, especially, by two-body-relaxation-driven mass
loss from clusters. Therefore, it is responsible for the evolution of the
cluster mass function. Moreover, the nuclear lifetime does not depend on cluster
mass, whereas the dynamical lifetime of a cluster is strongly related to its
mass. These differences cause different schemes in the evolution of the
luminosity and mass functions. Whereas the cluster luminosity function evolves
in coordination with the evolution of the most massive cluster members, the mass
function evolves faster at its low-mass end where the dynamical time is short.

Since the cluster luminosity function shows a more complicated evolution pattern,
we start our consideration with the CMF.

\begin{figure}
\resizebox{\hsize}{!}{
\includegraphics{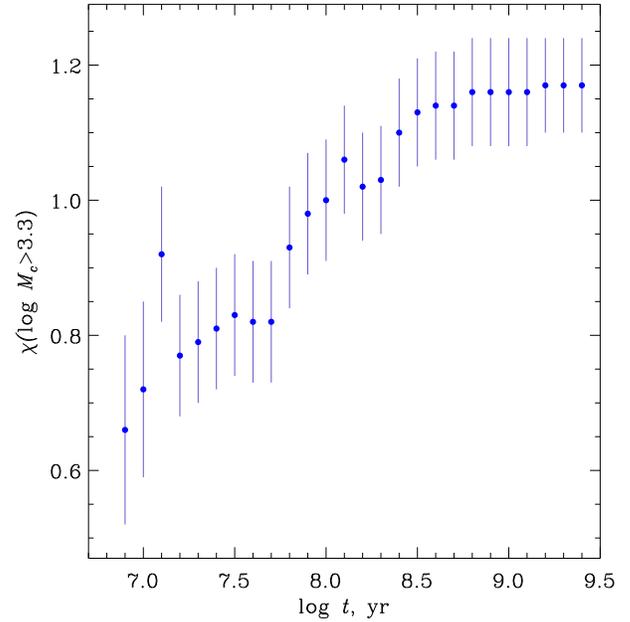}
}
\caption{Evolution of the slope $\chi$ of the mass function $\eta_t(\log M_c)$
at $\log M_c> 3.3$. The bars are Poisson errors.
}
\label{fig:sloev}
\end{figure}

\subsection{Evolution of the mass function of star clusters}

In Fig.~\ref{fig:fuma} we show mass spectra of three cluster subsamples
differing just by the upper limit of ages of the clusters. The first sample
contains the 49 youngest clusters with ages less than $t_y\approx8$ Myr ($\log
t\leqslant 6.9$). This is the lowest age limit where we have been able to
construct a significant mass function. Hereafter, we call this subset the
youngest sample. The second subsample includes additional moderately young
clusters and contains 207 entities ($\log t\leqslant7.9$). Finally, we consider
all 440 clusters ($\log t\leqslant9.5$). To improve the stability of
the solutions, we made re-binning on the mass scale requiring at least five
clusters per bin, but keeping the step of $\Delta\log M_c=0.15$ whenever
possible.

All three distributions demonstrate similar general patterns that can be
separated into two parts. The more massive part at $\log M_c>\log
M^*_c\approx3.3$ is almost linear on the logarithmic scales $\log \eta_t$ and
$\log M_c$ although the slope of the relation depends on the upper age limit. At
lower masses, the CPDMF changes into a non-linear pattern. After reaching the
maximum, the mass function of each sample starts to decrease. For the youngest
sample, the low mass portion of the CMF is rather flat and the turnover is less
significant. For older clusters, the CMF maximum is shifted to lower masses.
Choosing other subsamples of clusters by varying the upper age limit from $\log t
= 6.9$ to $\log t = 9.5$, we found that, at the age limit $\log t = 7.4$, the
CMF reaches a maximum at $\log M_c \approx 2$. If we add older and older clusters, 
this location does not change anymore, although the value of the CMF-maximum increases.

On the abscissa of Fig.~\ref{fig:fuma}, we observe different mass ranges for the
different subsamples. The lower mass limit depends significantly on the cluster
age. The younger the clusters, the higher is the lower mass limit
($M_c^{min}\approx 50 M_\odot$ at $\log t\leqslant 6.9$ and $M_c^{min}\simeq 5
M_\odot$ at $\log t\leqslant 9.5$). We consider this as a manifestation of mass
loss driven by the evolution of star clusters. On the other hand, the position
of the bin of the highest masses seems to be independent of the sample
considered. It varies from sample to sample only randomly, simply due to low
number statistics of high-mass clusters.

In Fig.~\ref{fig:sloev} we show the evolution of the slope of the linear portion
of the CMF. Again, one can split its evolutionary path into two parts. The
younger clusters show a steadily increasing slope from $\chi\approx 0.65$ at the upper limit
$t\approx 8$ Myr to $\chi\approx1.15$ at $t\approx 600$ Myr. At older ages, the
CMF slope practically does not change.

\begin{table}
\caption{Parameters of the CMF for the youngest sample.
}\label{tab:fitprm}
\begin{tabular}{cccrr}
\hline \noalign{\smallskip}
Fit & Segment & $\log M_c$&\mc{1}{c}{$\chi$}&\mc{1}{c}{$\log\eta^*$} \\
           & & range & & \\
\hline\noalign{\smallskip}
   S & low mass &$1.70\dots3.30^{1)}$ &$-0.18\pm0.14$&$-0.69\pm0.38$
\\
           & high mass&$3.30^{1)}\dots4.90$ &$0.66\pm0.14$&$2.16\pm0.56$ \\
\hline\noalign{\smallskip}
           & low mass &$1.70\dots3.37^{2)}$ &$-0.18\pm0.14$& \\
   C & & & &$-0.07\pm0.60$ \\
           & high mass&$3.37^{2)}\dots4.93$ &$0.66\pm0.14$ & \\
\hline\noalign{\smallskip}
\end{tabular}
$^{1)}$ $\log {M^*_c} = 3.30$\\
$^{2)}$ $log {M^*_c}$ from the solution of the continuity equation
\end{table}

The mass functions $\eta_t$ in Fig.~\ref{fig:fuma} and their slopes $\chi$ in Fig.~\ref{fig:sloev} describe cumulative mass distributions observed for clusters younger than $t$. Therefore, an increase in $\chi$ with increasing $t$ is a consequence of adding
older clusters, the mass spectra of which are similar or slightly flatter than the CIMF. At the limit, the slope of $\eta_t$ corresponds to the slope of the CPDMF representing a mixture of clusters of different ages.

\subsection{Initial mass function of star clusters}\label{sec:cimf}

In the youngest sample, even the least massive clusters seem to have not had enough 
time to evolve dynamically. Therefore, we assume in the following that
the youngest sample gives a good approximation of the CIMF, and the youngest
point of the evolution curve in Fig.~\ref{fig:sloev} is close to the slope of
the CIMF. According to Fig.~\ref{fig:fuma}, we assume that Galactic clusters do
form in a mass range from $M_c^{min}=50\,M_\odot$ to
$M_c^{max}=3\cdot10^5\,M_\odot$.

The two segments of their CMF ($\log M_c < \log M^*_c \approx 3.3$, and $\log
M_c > \log M^*_c$) can be scaled by a power law function with
Eq.~(\ref{eq:eta}). The corresponding parameters are given in
Table~\ref{tab:fitprm} (Fit S) where the fits are carried out separately for
each of the two CMF segments. Again, since the CMF is based on clusters within
the completeness limit, we consider the flattening $\eta_t(M_c)$ at $\log M_c <
3.3$ to be a real feature and not a signature of incompleteness.

To convert Eq.~(\ref{eq:eta}) in a convenient form suitable for the
complete range of cluster masses, we adjust the solutions making use of the CMF
continuity condition, so the mass $M^*_c$ can be determined where for both
segments $\eta_t(M^*_c) = \eta^*$. The results of these fits are given in
Table~\ref{tab:fitprm} (Fit C). Obviously, the transformations do not affect the
CMF slopes $\chi$. Applying these definitions, we obtain
\begin{equation}
\eta_t = \frac{dN}{d\log M_c} = \eta^* \, \left(\frac{M_c}{M^*_c}\right)^{-\chi}.
\label{eq:eta2s}
\end{equation}

For the construction of the CIMF we introduce a birth function $\beta(M_c,t)$
of star clusters as
\[
\frac{\partial ^2N}{\partial M_c\partial t} = \beta(M_c,t).
\]
Provided that the mass distribution of clusters at birth (CIMF) does not depend
on time, we can factorize to get
\[
\beta(M_c,t) = \zeta(M_c)\,\upsilon(t)\,,
\]
where $\zeta(M_c)$ is the CIMF, and $\upsilon(t)$ is the cluster formation rate
(CFR). Provided that at $t>t_0$ the violent events related to infant
mortality \citep[see e.g. ][]{kah01} have generally ceased and a cluster starts
its post-natal life, one can assume that within a short time interval following
this moment, i.e. between $t_0=4$ Myr and $t_y=8$ Myr, the mass loss from a
cluster due to two-body relaxation seems to be insignificant \citep[][]{lamea},
and the relation between CMF and CIMF is straightforward:
\[
\zeta(M_c)= \frac{\log e}{\bar{\upsilon}(t)\Delta t}\,\frac{\eta_t}{M_c}\,,
\]
where $\bar{\upsilon}(t)$ is the average cluster formation rate during the time
$\Delta t$. Using Eq.~(\ref{eq:eta2s}) and the normalization condition
\[
\int\limits_{M_c^{min}}^{M_c^{max}}\zeta(M_c)\,dM_c = 1\,, 
\]
we derive
\begin{equation}\label{eq:cimf}
\zeta(M_c)=\frac{k}{M^*_c}\,\left(\frac{M_c}{M^*_c}\right)^{-\alpha}\,,
\end{equation}
where $\alpha=1+\chi$, and
\begin{equation}\label{eq:kfac}
k=\frac{\log e\,\eta^*}{\bar{\upsilon}(t)\Delta t}
\end{equation}
is the normalization factor. With the parameters from Table~\ref{tab:fitprm} one
gets
\[
\alpha = \left\lbrace
\begin{array}{lll}
0.82\pm0.14, & \textrm{ for } M_c^{min} &< M_c < M^*_c \\
1.66\pm0.14, & \textrm{ for } M^*_c &< M_c < M_c^{max},
\end{array}
\right.
\]
and $M^*_c=2.3\cdot10^3\,M_\odot$. The corresponding normalization factor is
$k=0.24$.

The two-segment CIMF can be understood in terms of the infant mortality process.
The more massive clusters being able to survive the infant mortality and to keep
the memory of their parental conditions in the embedded phase show a power law
distribution scaling with their natal distribution. This interpretation is
supported by the results of \citet{lbs91}, who studied dense cores within the
molecular cloud L~1630 and found that the mass distribution of dense cores can
be fitted by a power low with an exponent of $-1.6$. The result is in remarkable
agreement with $-\alpha = -1.66$ we obtained for the CIMF at $M_c >
2.3\cdot10^3\,M_\odot$.

The low-mass segment of the CIMF probably consists for
the major part of remnants of clusters damaged during the infant mortality phase
and therefore lost memory of their recent past. Their mass distribution in the
logarithmic scale resembles rather white noise. On the other hand, among young
low-mass clusters, there may be a fraction that already had low masses at birth.
Although the number of remnants and originally low-mass clusters is relatively
large, their proportion is considerably less than the one they had if they
would follow the power law of the more massive clusters.

Using the conclusions of a recent study of the influence of the
gas expulsion phase on the shape of the resulting CIMF \citep{parea08}, we infer that the empirical CIMF derived in the present paper is compatible with
the scenario of a low, randomly distributed star formation efficiency with an
average of 20\% and a dispersion of 3\%.

\begin{figure}
\resizebox{\hsize}{!}{
\includegraphics{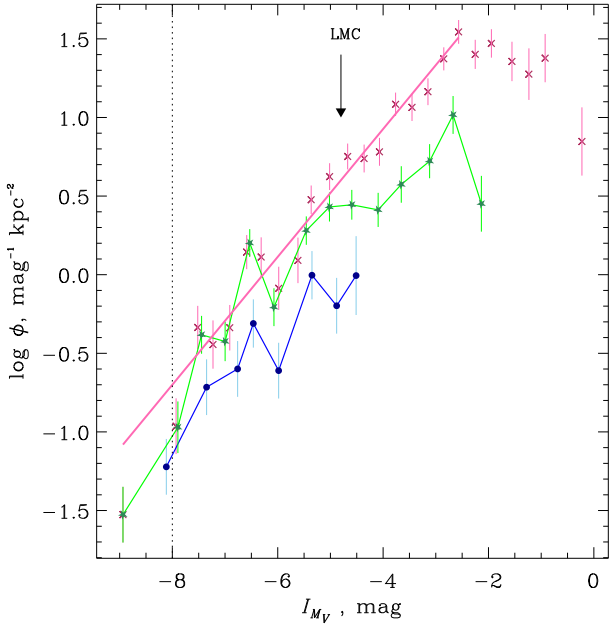}
}
\caption{Evolution of the luminosity function $\phi(I_{M_V})$ of Galactic open
clusters. Different symbols mark samples with different upper limits of cluster
age. The filled circles are for clusters with $\log t \leqslant 6.9$, stars
show CLF for $\log t\leqslant 7.9$, and crosses indicate the CPDLF based on all
440 clusters ($\log t \leqslant 9.5$). The bars are Poisson errors. The arrow
indicates the observed limit of integrated absolute magnitudes reached for open
clusters in the LMC. The dotted vertical line at $I_{M_V} = -8$ separates the
luminous part where the CLF deficiency was observed for extragalactic clusters
by \citet{slars}.
}
\label{fig:fulu}
\end{figure}

\subsection{Evolution of luminosity function of star clusters}

The evolution of the luminosity function of Galactic star clusters is shown in
Fig.~\ref{fig:fulu}. Similar to the mass function, the CLF apparently includes
two regimes (although less clearly distinguishable from each other than in the
CMF case). The dominant feature is the linear regime of the CLF, occupying the
bulk of brighter magnitudes up to $I_{M_V} \approx -2.5$, and a minor detail is
a tail at fainter magnitudes. However, since the evolution of the luminosity
function is governed mainly by stellar evolution, and only at the end of cluster
life by dynamical effects, there is no direct correspondence between absolute
magnitude and mass of clusters. In most of the CLF bins of absolute magnitude,
we find a mixture of masses and evolutionary states. Only at the extremes of the
magnitude scale, there are clusters of more or less ``pure'' evolutionary
status. As a result, the initial luminosity function of open clusters can be taken from
the distribution of the youngest clusters ($\log t \leqslant6.9$) at
$I_{M_V}<-5.5$ where the clusters belong to the massive portion of the CIMF
($\log M_c>3.3$), and the tail of the CPDLF at $I_{M_V}>-2$ consists mainly of
decaying clusters with masses less than $\approx 100 M_\odot$ i.e., after the
CPDMF maximum.

Contrary to the CMF, the luminosity function at $-8.0<I_{M_V}<-5.5$ does not show
any evolution of the slope with time. It seems that, with increasing age, the CLF
is just moving along the ordinate due to the arrival of newborn clusters
without changing the shape and extends to fainter magnitudes due to the aging
effect (fading of the brightest stars).

\section{Discussion of the results}\label{sec:discuss}

Based on this study on local Galactic clusters, we find that, in the immediate
vicinity of the Sun, young ($\log t \leqslant $6.9) open clusters are present
within a wide range of masses. The mass range extends from the highest masses -
observed also in other galaxies - to a few tens of solar mass. This is a strong
hint that the population in the solar neighbourhood is typical of spiral
galaxies. In quite a natural way, cluster-like associations (see. Sect.~\ref{sec:dadef}) form the high-mass end of
the CPDMF; in other words, they represent the same population of the Galactic
disc as young open clusters. A comparison of our luminosity and mass functions
of star clusters with those in other galaxies, which we discussed in
Sect.\ref{sec:lmf}, strongly indicates a universal law of cluster
formation, which is only weakly dependent on the environment. The issue of the
universality of the cluster initial mass function in different galaxies is
discussed in more detail by \citet{grijs03}.

Studying the CPDLF of extragalactic clusters, \citet{slars} notes a depletion of
the CPDLF at $I_{M_V}<-8$ in three galaxies. The author explains this effect by
a possible increase in the steepness of the CPDLF. {\citet{giel06} confirm the
existence of this precipitous drop in the cluster luminosity functions of the
spiral galaxies NGC~6946 and M51 and explain this by assuming two-segment-, or
Schechter CIMFs, in these galaxies. We observe the same feature in the luminosity
function of Galactic clusters, too. Unfortunately, the \ascc sample is
restricted to the solar neighbourhood and does not include enough
of the brightest clusters to be compared with the brightest limit of
the extragalactic CPDLFs. As noted at the end of the last section, the parallel
slope of the CPDLF for samples of different ages in the range $-8 \leq I_{M_V}
\leq -5$ is a consequence of the continuity equation. A depletion at
$I_{M_V}<-8$ suggests that no open clusters are born (or leave their parental
cloud) with an absolute integrated magnitude much brighter than $I_{M_V}=-8$. A
caveat is the poor statistics (only a few extragalactic clusters and no Galactic
ones).

From the physical point of view, the observed deficiency might correspond
to a gap caused by so-called ``type II embedded clusters'' \citep{krobo02},
which are massive enough to produce plenty of O stars providing fast gas
expulsion and parent cluster decay. The low-mass limit of this cluster family
should be in the vicinity of $\log M_c=4.5$ in this case.

\citet{lala03} have compiled a catalogue of about 100 embedded clusters within 2.4
kpc from the Sun. The sample contains some optical objects and is partly
overlapping with our data. Using models of the luminosity function,
\citet{lala03} scaled the IR counts within the areas studied, estimated cluster
masses, and constructed a mass function of embedded clusters. Typically, the
clusters are distributed over a mass range from 50 to 1000~$M_\odot$ and thus,
related the low-mass part of our CIMF. Based on ``reasonably complete
subsamples'', the mass distribution function $M_c\,dN/d\log M_c$ increases
approximately linear with $M_c$. This indicates that the CMF of embedded
clusters $dN/d\log M_c$ is nearly constant and agrees with
low-mass CIMF (see Table~\ref{tab:fitprm}).

The newly determined CIMF and CPDMF have major implications for the evolution of
the Galactic disc population. The average mass of forming clusters can be
calculated from Eq.~(\ref{eq:cimf}):
\[
\olin{M}^{CIMF}_c = \int\limits_{M_c^{min}}^{M_c^{max}}M_c\,\zeta(M_c)\,dM_c\,.
\]
With the parameters determined in Sect.\ref{sec:cimf}, we find that $
\olin{M}^{CIMF}_c = 4.5\cdot 10^3\,M_\odot\,$. The accuracy of this estimate
depends on the accuracy of the CIMF parameters and, in particular, on the
adopted integration limits. For example, by excluding the highest mass bin
(i.e., a shift of $M_c^{max}$ from $3\cdot10^5\,M_\odot$ to to $10^4\,M_\odot$)
from consideration, one obtains $\olin{M}^{CIMF}_c = 1.4\cdot 10^3\,M_\odot\,$.

The average mass of the presently observed clusters $\olin{M}^{CPDMF}_c$ can be
found from the CPDMF by its integration as
\[
\olin{M}^{CPDMF}_c = \frac{\int\limits_{\log M_c^{min}}^{\log
M_c^{max}}M_c\,\eta(M_c)\,d\log M_c}
{\int\limits_{\log M_c^{min}}^{\log M_c^{max}}\eta(M_c)\,d\log M_c}\,.
\]
Taking the mass limits and the CPDMF from Fig.~\ref{fig:fuma}, we get
$\olin{M}^{CPDMF}_c \simeq 700\,M_\odot$. The accuracy of this value is
estimated to be about 30\%. The value of $\olin{M}^{CPDMF}_c$ coincides
perfectly with what is usually assumed for open clusters. The average age of the
presently observed clusters computed in the same way is equal to about 260~Myr.
We stress that this estimate is based on purely empirical grounds. A comparison
of $\olin{M}^{CIMF}_c$ and $\olin{M}^{CPDMF}_c$ indicates that, during the first
260~Myr of its evolution, a typical open cluster loses about 60\%--80\% of its
initial mass. We consider this as evidence of the importance of this phase
in the evolution of open clusters. The average mass-loss rate based on these
numbers then ranges from 3 to 14~$M_\odot$ Myr$^{-1}$. Since this estimate
follows from observed masses of clusters, it includes both, mass loss due to
stellar evolution and dynamical ejection of cluster stars.

The average surface density of open clusters in the solar neighbourhood $n_c$
can be computed as
\[
n_c = \int\limits_{\log M_c^{min}}^{\log M_c^{max}}\eta(M_c)\,d\log M_c \approx
86\,\, \mathrm{kpc}^{-2}\,.
\]
This is about 25\% lower than was estimated in our study of the cluster
population \citep[][]{clupop} on the basis of a volume-limited sample of local
clusters closer than 850 pc. This disagreement can be explained by our omitting
of clusters with $\hat{I}_V>8$ in the present estimate. When taken into
account, the estimates differ only by about 10\%, which may result from the
different methods of the sample extraction, i.e. the construction of
volume-limited versus magnitude-limited samples.

Let us estimate the input of open clusters into the Galactic disc population and
consider the total mass of optical star clusters that were formed during the
history of the Galactic disc. The surface density of the Galactic disc stars
having passed through an open cluster phase can be written as
\[
\Sigma = \bar{\upsilon}\,T\,\olin{M}^{CIMF}_c\,,
\]
where $T=13$ Gyr is the age of the Galactic disc, and
$\bar{\upsilon}=0.4\,\,\mathrm{kpc}^{-2}\mathrm{Myr}^{-1}$ is the CFR determined
via Eq.~(\ref{eq:kfac}). This value is determined as an average over the past 4
Myr and has to be compared with other determinations. For optical clusters, we
find similar results in the literature: e.g.
$0.45\,\,\mathrm{kpc}^{-2}\mathrm{Myr}^{-1}$ in \citet{bcd91}, and
$0.25\,\,\mathrm{kpc}^{-2}\mathrm{Myr}^{-1}$ in \citet{ecle85}. For embedded
clusters, a much higher formation rate of
$2-4\,\mathrm{kpc}^{-2}\mathrm{Myr}^{-1}$ is obtained by \citet{lala03}. This
may be regarded as evidence of a considerable infant mortality among
embedded clusters.

Assuming that the recent average CFR has not changed during the history of the
Galactic disc one finds that for $\olin{M}^{CIMF}_c=4.5\cdot 10^3\,M_\odot$
\[
\Sigma = 22\,M_\odot\,\textrm{pc}^{-2}\,.
\]
This amount corresponds to about 40\% of the total surface density of the
Galactic disc in the solar neighbourhood that, according to \citet{holmfly04}\
is $56\pm6\, M_\odot\,\textrm{pc}^{-2}$. From data on embedded clusters,
\citet{lala03} find $\Sigma = 10-30\,M_\odot\,\textrm{pc}^{-2}$. The value of
$\Sigma = 22\,M_\odot\,\textrm{pc}^{-2}$ obtained here is considerably higher
than the previous estimates for the input of open clusters to the observed
stellar population of the Galactic disc that is quoted as about 10\%
\citep[see][]{misca78,clupop} or even less than 10\% \citep{wiel71}. These
rather low estimates can be explained by several situations: they are based either
on insufficient cluster statistics or on cluster masses from underestimated
stellar counts, and on uncertain virial masses of open clusters or on
arbitrarily chosen estimates of typical cluster contents. The newly derived
value of $\Sigma$ is in agreement with indirect evidence of binary
population statistics \citep[see][]{krok195}. This value significantly
readdresses the problem of the main sources of the Galactic disc population and
indicates open clusters to be one of the major factories of field stars, in line
with the fact that a major fraction of star formation seems to occur in embedded
clusters \citep[][]{zmw93,lala03,megea05}.

\section{Summary and conclusions}\label{sec:concl}

Using the integrated magnitudes computed from individual photometry of cluster
members available in the \ascc, we found that the sample of 648 clusters drawn
from this catalogue is complete for apparent integrated magnitudes brighter than
8 mag, with 440 clusters above this completeness limit. The corresponding
completeness distances depending on the brightness of clusters are typically
between 1 and 3 kpc, although some prominent clusters can be seen up to
distances of 10 kpc and more.

The CPDLF constructed on the basis of the complete sample is observed in the
range of $I_{M_V}= [-10,\,-0.5]$ mag, i.e. about 5 magnitudes deeper than
observable in nearby galaxies. It increases from the brightest limit to a
turnover at about $I_{M_V}\approx-2.5$ and slowly falls down towards fainter
magnitudes. We consider the turnover to be a real feature reflecting the
behaviour of Galactic open clusters. At magnitudes brighter than the turnover
position, the CPDLF is linear with a slope $a=0.41\pm0.01$ (or
$\alpha=2.02\pm0.02$), which is in perfect agreement with the observed slopes of
cluster luminosity functions in nearby spiral galaxies, although these show a
higher star formation activity.

The masses of Galactic open clusters cover a range from several tens to hundreds
of thousands of solar masses. The CPDMF mimics the general behaviour of the
CPDLF: it shows a linear part at high masses ($\log M_c > 2.5$), a broad maximum
between $\log M_c=1.5$ and 2.5, and a slow decline towards lower masses. In
spite of the present-day observations only allowing for a comparison
of the CPDMF data of extragalactic open clusters drawn from model
mass-luminosity relations, the agreement with published values is impressive. We
find that for $\log M_c >2.5$ the CPDMF can by fitted with high confidence by a
power law with a slope $\chi=1.01\pm0.04$. We regard the observed coincidence of
cluster mass function parameters derived for the Galactic and extragalactic
clusters with different methods as indirect evidence that both methods of
estimating cluster masses are calibrated consistently and give correct and
reliable results.

Inspection of cluster samples with different upper limits for age indicates that
cluster mass spectra change with time. At every age the cluster mass function
keeps the basic features of the CPDMF, i.e., a quasi-linear high-mass portion,
and a non-linear portion at lower masses. With time, the slope of the linear
portion increases from $\chi=0.66\pm0.14$ at $\log t\leqslant6.9$ to
$\chi=1.13\pm0.08$ at $\log t\leqslant8.5$. Then the steepening of the CPDMF
slows down and arrives at $\chi=1.17\pm0.07$ for $\log t\leqslant9.5$. The
low-mass portion changes from an approximately flat distribution at $\log t=6.9$
to a clearly non-linear behaviour displaying a broad maximum with a peak at
about 100 $M_\odot$.

We construct the
CIMF from the data of a subsample of the 49 youngest clusters with $\log t
\leqslant6.9$. The CIMF has a segmented structure with two power-law segments:
the power-law slopes are $\alpha=1.66\pm0.14$ for $\log M_c=3.37\dots4.93$ and
$\alpha=0.82\pm0.14$ for $\log M_c=1.7\dots3.37$.

The luminosity function of open clusters does not show the same systematic
steepening with age as does the mass function. It evolves via parallel-shifting
along the ordinate without changing the slope. Only the faintest segment of the
CLF seems to depend on age and spreads towards the fainter magnitudes with age.
The weak dependence on age and the presence of a mixture of clusters with
different masses in every magnitude bin makes it difficult to determine the CIMF
from the CLF.

When comparing the average mass of the newly formed, youngest clusters
$\olin{M}^{CIMF}_c\simeq 4.5\cdot 10^3\,M_\odot$ with the average cluster mass
from the whole sample ($\olin{M}^{CPDMF}_c\simeq 700\,M_\odot$), one finds, as
expected, that clusters are typically much more massive at birth than later.
Since this observation concerns ``normal'' optical clusters, one should
acknowledge the importance of a continuous mass loss occurring in open clusters
during their evolution with a rate around 3-14 $M_\odot$ Myr$^{-1}$.

If this effect is not taken into account, one underestimates the role that open
clusters play in supplying the Galactic disc with the products of their
evolution: stars and stellar remnants, sub-stellar objects, gas and dust.
Provided that the cluster formation history has not changed dramatically in the
solar neighbourhood during the evolution of the Galactic disc, at least 40\% of
the observed surface density of the disc comes from open clusters. The other
half is left for violent events (strong winds, HII zones, or supernova
explosions) provoking the fast mass loss during the ``infant'' phase of cluster
evolution proposed first by \citet{tut78} and later by \citet{kah01}, and
recently considered in detail by \citet{baukr07} and \citet{weidea07}.

\begin{acknowledgements}
We thank the anonymous referee for his/her useful comments.
This study was supported by DFG grant 436 RUS 113/757/0-2, and RFBR grants
06-02-16379, 07-02-91566.
\end{acknowledgements}

\bibliographystyle{aa}
\bibliography{clubib}

\end{document}